\documentclass[twoside]{article}
\usepackage{spconf,amssymb,amsmath,epsfig}
\usepackage{multirow}
\usepackage{hyperref}
\ninept  % 9-pt size optional (10-pt default)

\title{A Study on Zero-Shot Non-Intrusive Speech Intelligibility for Hearing Aids Using Large Language Models}
\name{Ryandhimas E. Zezario$^1$, Dyah A.M.G. Wisnu$^{12}$, Hsin-Min Wang$^1$,Yu Tsao$^1$}
\address{
  $^1$Academia Sinica 
  $^2$National Chengchi University
  }

\begin{document}
\maketitle
\begin{abstract}
This work focuses on zero-shot non-intrusive speech assessment for hearing aids (HA) using large language models (LLMs). Specifically, we introduce GPT-Whisper-HA, an extension of GPT-Whisper, a zero-shot non-intrusive speech assessment model based on LLMs. GPT-Whisper-HA is designed for speech assessment for HA, incorporating MSBG hearing loss and NAL-R simulations to process audio input based on each individual's audiogram, two automatic speech recognition (ASR) modules for audio-to-text representation, and GPT-4o to predict two corresponding scores, followed by score averaging for the final estimated score. Experimental results indicate that GPT-Whisper-HA achieves a 2.59\% relative root mean square error (RMSE) improvement over GPT-Whisper, confirming the potential of LLMs for zero-shot speech assessment in predicting subjective intelligibility for HA users.

\end{abstract}
\section{Introduction}
Assessing speech intelligibility is a crucial factor in evaluating hearing aid (HA) applications \cite{barker22_interspeech}. The most direct approach is to have human listeners recognize words from an audio sample. However, despite their reliability, human-based assessments are costly and impractical for unbiased evaluations, as a sufficient number of listeners is required to minimize bias.

With two series of Clarity Prediction Challenges \cite{barker22_interspeech, 10446441}, there is growing interest in developing non-intrusive speech intelligibility prediction models for HA. However, despite notable achievements, a major challenge in deploying reliable non-intrusive speech intelligibility models is the availability of sufficient training data, as data collection can be time-consuming and labor-intensive. Recently, with the growing popularity of large language models (LLM), studies have explored their potential for speech assessment. One such approach is GPT-Whisper \cite{zezario2025studyzeroshotnonintrusivespeech}, which consists of an audio-to-text module and GPT-4o for evaluating the predicted word sequence to obtain estimated scores. Experimental results confirm that GPT-Whisper demonstrates a moderate correlation with intelligibility metrics. Furthermore, considering its notable performance and the need for a reliable non-intrusive speech intelligibility model that can be deployed in a zero-shot setting, we aim to extend GPT-Whisper for zero-shot non-intrusive speech intelligibility assessment in HA.

In this study, we propose an extension of GPT-Whisper for HA, namely GPT-Whisper-HA. GPT-Whisper-HA is specifically designed for speech assessment in hearing aids by incorporating MSBG hearing loss and NAL-R simulations to mimic the audio input based on an individual's hearing profile. Furthermore, two ASR modules are selected as judges to assist in determining whether the input audio is easily recognizable. Unlike typical audio data, hearing-impaired speech may not be easily recognized by ASR systems. We assume that if both a simpler and a more advanced ASR module can easily recognize the word sequence, the audio input is likely clear. Conversely, if the two ASR modules show differing trends, the audio may contain noise or distortions that reduce intelligibility. GPT-4o is then used to generate the assessment score, leveraging naturalness as the key metric, following the original GPT-Whisper framework. Finally, score averaging is performed to obtain the final estimated score.

\section{GPT-Whisper-HA}
\begin{figure}[t]
\centering
\includegraphics[width=6 cm]{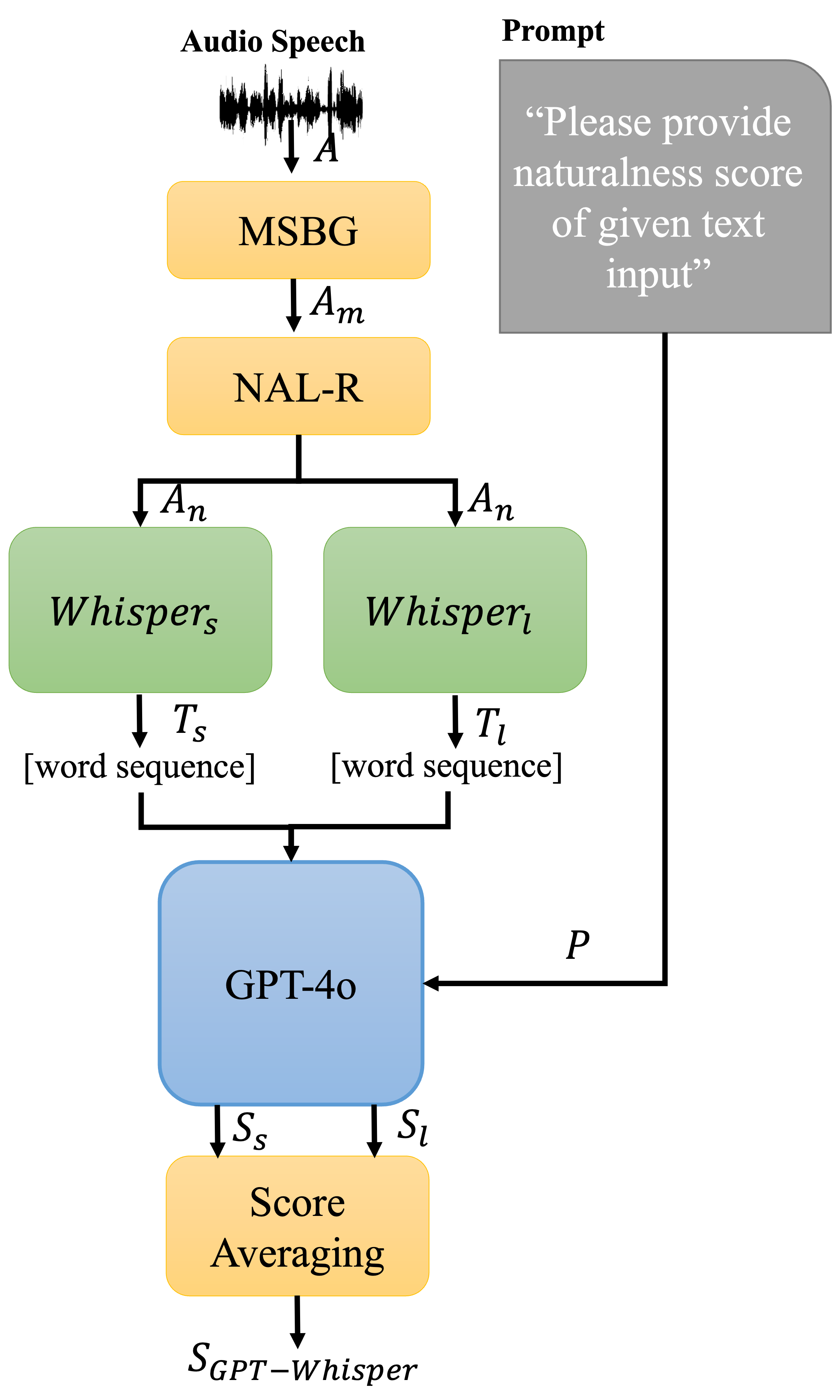} 
\caption{Zero-shot speech intelligibility for HA with GPT-Whisper-HA.} 
\label{fig:gptwhisper}
\end{figure}

The overall framework of GPT-Whisper-HA is shown in Fig. \ref{fig:gptwhisper}, where we selected two variants of Whisper \cite{Whisper}—small ($Whisper_{s}$) and large ($Whisper_{l}$)—as the audio-to-text modules due to their robust performance in accurately transcribing speech across diverse acoustic conditions,

\graphicspath{ {./images/} }

Specifically, given an input audio $\textit{A}$, the audio input is first processed by MSBG and NAL-R before undergoing audio-to-text conversion using Whisper ASR, as defined below:

\begin{equation}
\label{eq:audiototext}
   \small
    \begin{array}{c}
    A_{m} = MSBG(A),
    \\
    A_{n} = NALR(A_{m}),
    \\
    T_{s}= Whisper_{s} (A_{n}),
    \\
    T_{l}= Whisper_{l} (A_{n}).
    \\
    \end{array} 
\end{equation}
Based on the predicted word sequences $T_{s}$ and $T_{l}$, we use GPT-4o to estimate two assessment scores, $S_{s}$ and $S_{l}$. For prompt $P$ engineering, we assess the text representation based on the naturalness score, which measures how similar the predicted text is to human-generated text in terms of fluency, coherence, and context. The final GPT-Whisper score is obtained through score averaging, with the detailed process defined as follows.

\begin{equation}
\label{eq:gptwhisper}
   \small
    \begin{array}{c}
    S_{s}= GPT4o(T_s,P),
    \\
    S_{l}= GPT4o(T_l,P),  
    \\
    S_{GPT-{whisper}}= ScoreAve(S_s,S_l),     
    \end{array} 
\end{equation}
\section{Experiments}
\subsection{Experimental Setup}
The Clarity Prediction Challenge (CPC) 2023 dataset \cite{10446441} consists of recordings from six talkers and ten enhancement methods, each representing a different HA system, with corresponding subjective intelligibility scores for the output of each HA. Specifically, we select Track 1 of the test set, which contains 305 utterances, to evaluate our system. The evaluation metrics include root mean square error (RMSE), linear correlation coefficient (LCC), and Spearman’s rank correlation coefficient (SRCC). A lower RMSE signifies better alignment with ground-truth scores, while higher LCC and SRCC values indicate stronger correlations between predictions and actual scores. All supervised models in this study were trained on the CPC 2023 training set for fair comparison.

\subsection{Experimental Results}
To evaluate the performance of GPT-Whisper-HA, we prepared three system comparisons: MBI-Net \cite{zezario2022mbi}, MBI-Net+ \cite{zezario24_interspeech}, and GPT-Whisper \cite{zezario2025studyzeroshotnonintrusivespeech}. MBI-Net and MBI-Net+ achieved top-three performances among the best non-intrusive systems in the first and second Clarity Challenges. Both models were trained using the Track 1 training set of the CPC 2023 dataset. For model architecture, both MBI-Net and MBI-Net+ employ convolutional layers with channel sizes of 16, 32, 64, and 128, a one-layer Bidirectional Long Short-Term Memory (BLSTM) network with 128 nodes, a fully connected layer with 128 neurons, and an attention mechanism. The key difference between the models is that MBI-Net+ utilizes Whisper to extract cross-domain features, while MBI-Net uses WavLM. Additionally, MBI-Net+ incorporates additional modules for objective-based assessment metrics and a classifier module to distinguish different HA model inputs. For GPT-Whisper, we follow the original setup to predict the estimated score.

Table 1 presents the evaluation results of different models in terms of LCC, SRCC, and RMSE. Among the models, MBI-Net+ achieves the highest performance, with an LCC of 0.721, an SRCC of 0.714, and the lowest RMSE of 28.370. MBI-Net, while slightly behind MBI-Net+, also demonstrates competitive performance with an LCC of 0.669, an SRCC of 0.665, and an RMSE of 30.260. These results align with our assumption, as both models benefit from supervised training, allowing them to optimize their predictions using labeled data. Interestingly, GPT-Whisper-based models, which operate in an unsupervised manner, exhibit moderate correlation scores for LCC (0.541) and SRCC (0.501). Despite their lower prediction performance, the predicted scores from GPT-Whisper still show some moderate correlation with subjective human ratings. Notably, our proposed GPT-Whisper-HA, which incorporates HA-related adaptations into a zero-shot modeling strategy, enhances overall prediction performance—improving LCC from 0.541 to 0.570, SRCC from 0.501 to 0.558, and reducing RMSE from 37.019 to 34.767. Furthermore, employing two ASR models as judges for the final assessment score demonstrates consistent improvement in prediction performance in the zero-shot scenario.

\begin{table}[t]
\caption{LCC, SRCC, and RMSE results of GPT-Whipser-HAS and other methods.}
\footnotesize
\begin{center}
 \begin{tabular}{c||c||c||c||c} 
 \hline
 \hline
 \textbf{Model} &Unsupervised &\textbf{LCC} & \textbf{SRCC} & \textbf{RMSE}  \\ [0.5ex] \cline{2-5}
 \hline\hline

MBI-Net \cite{zezario2022mbi} &No&0.669&0.665&30.260\\\hline
MBI-Net+ \cite{zezario24_interspeech}&No&\textbf{0.721}&\textbf{0.714}&\textbf{28.370}\\\hline
GPT-Whisper \cite{zezario2025studyzeroshotnonintrusivespeech}&Yes&0.541&0.501&37.019\\\hline
GPT-Whisper-HA&Yes&0.570&0.558&34.767\\\hline
 \hline
\end{tabular}
\end{center}
\end{table}

\section{Conclusions}

In this paper, we proposed GPT-Whisper-HA, a zero-shot model that incorporates HA-related adaptations for speech intelligibility prediction. Our results demonstrate that GPT-Whisper-HA outperforms the baseline GPT-Whisper model, improving LCC from 0.541 to 0.570, SRCC from 0.501 to 0.558, and reducing RMSE from 37.019 to 34.767. Additionally, employing two ASR models as judges for the final assessment further enhances prediction performance. These improvements highlight the effectiveness of integrating hearing aid-specific features into a zero-shot speech intelligibility model.

\bibliographystyle{IEEEbib}
\bibliography{refs}
\end{document}